\documentclass[12pt,preprint]{aastex}

\begin{document}

\title{Near-Infrared Photometry of four stellar clusters in the Small 
Magellanic Cloud
\footnote{Based on observations collected at La Silla ESO 
Observatory under proposal 076.D-0381(B).}}

\author{Alessio Mucciarelli}
\affil{Dipartimento di Astronomia, Universit\`a 
degli Studi di Bologna, Via Ranzani, 1 - 40127
Bologna, ITALY}
\email{alessio.mucciarelli@studio.unibo.it}

\author{Livia Origlia}
\affil{INAF - Osservatorio Astronomico di Bologna, Via Ranzani, 1 - 40127
Bologna, ITALY}
\email{livia.origlia@oabo.inaf.it}

\author{Claudia Maraston}
\affil{Institute of Cosmology and Gravitation, 
University of Portsmouth, Mercantile House, Hampshire Terrace, 
P01 2EG, Portsmouth, United Kingdom}
\email{claudia.maraston@port.ac.uk}

\author{ Francesco R. Ferraro}
\affil{Dipartimento di Astronomia, Universit\`a 
degli Studi di Bologna, Via Ranzani, 1 - 40127
Bologna, ITALY}
\email{francesco.ferraro3@unibo.it}

\begin{abstract}  

We present high-quality 
J, H and $K_s$ photometry of four Small Magellanic Cloud stellar 
clusters with intermediate ages in the  
1--7 Gyr range (namely NGC~339, 361, 416 and 419)  .
We obtained deep Color-Magnitude Diagrams  
to study the evolved sequences and providing a detailed
census of the Red Giant Branch (RGB), Asymptotic Giant Branch (AGB) and Carbon 
star populations in each cluster and their contribution to the 
total cluster light. 
We find that in the $\sim$5-7 Gyr old clusters 
AGB stars account for $\sim$6\% of the total 
light in $K_s$-band,
Carbon stars are lacking 
and RGB stars account for $\sim$45\% of the total bolometric luminosity.
These empirical findings are in good agreement with the theoretical 
predictions. Finally, we derived photometric metallicities computed 
by using the properties of the RGB and finding an iron content of 
[Fe/H]=~-1.18, -1.08, -0.99 and -0.96 dex for  NGC~339, 361, 416 and 419
respectively.

\end{abstract}  
 

\section{Introduction}   
\label{intro}

The history and the evolution of the Large and Small Magellanic Clouds 
(LMC and SMC, respectively) are intimately related 
to the gravitational interactions 
between the two Clouds and the Milky Way. 
In particular, the main episodes of star formation enhancement in the SMC 
are triggered by the near, perigalactic passages of the LMC and of
the Milky Way \citep{hz04, zh04}. 
The epoch and rate of the main star formation episodes in the Magellanic Clouds (MCs) 
are still matter of debate and several scenarios have been proposed. 
For the SMC, while \citet{hz04} suggest 3 episodes occurred 400 Myr, 
3 Gyr and 9 Gyr ago, \citet{dolphin} 
favor a more continuous star formation in the halo with a dominant 
episode 5--8 Gyr ago. \citet{raf} argue that the cluster age 
distribution shows 
a few peaks, but no significant {\sl gaps} as in the LMC \citep[see also][]{chiosi06}.

MCs host a globular cluster system which includes objects with different age and metallicity, 
thus representing a formidable probe of 
the various stellar populations in the MCs as well as ideal templates for the 
study of stellar evolution and population synthesis.
In this respect, the near-infrared (IR) spectral range is particularly suitable 
to sample the evolved stellar
sequences, whose giant stars are characterized by low surface gravities 
and effective temperatures.

In our previous papers \citep[][hereafter Paper I and Paper II, respectively]{f04,m06} 
the near-IR Color-Magnitude Diagrams
(CMDs) of 19 young-intermediate age LMC clusters (from $\sim$80 Myr 
to $\sim$3 Gyr) have been analysed, providing a  
quantitative estimate of the 
AGB and RGB contributions to the total light, 
as a function of the cluster age.
The AGB contribution 
to the total luminosity starts to be significant at $\sim$200 Myr, 
with a maximum ($\sim$80\%) at $\sim$500-600 Gyr.  
At this same epoch 
the RGB phase transition occurs and for ages older than 1 Gyr the RGB itself becomes 
fully developed, while the contribution of AGB is progressively 
reduced.

The present paper reports the results for four SMC clusters  
belonging to the intermediate-age population of the SMC.
The principal aims of this 
work are the study of the main features of
their near IR CMDs and the contribution to the total cluster luminosity 
of the AGB and RGB stars. 
This sample of SMC clusters allows to check the contribution 
of the AGB and RGB stars for clusters in an age range ($\sim$5-7 Gyr) not 
covered by the LMC cluster system (because it corresponds to the so-called 
{\sl Age-Gap}), thus providing complementary information.

The paper is organized as follows: Sect.~\ref{obs} describes the observations and 
photometric analysis. Sect.~\ref{cmd} presents the near-IR CMDs 
and their main features,
the inferred metallicities and the integrated magnitudes.
Sect.~\ref{count} describes the procedure adopted to estimate 
the completeness correction and the field decontamination. 
Sect.~\ref{agb} describes the detailed census of 
the AGB and Carbon stars in each cluster, and their contribution to the total luminosity, 
while Sect.~\ref{rgb} analyzes the RGB stellar population.

\section{Observations and data reduction}
\label{obs}

A set of J, H and $K_{s}$ images of four stellar clusters 
(namely NGC~339, 361, 416 and 419) in the SMC 
has been selected at the European Southern Observatory 
(ESO), La Silla, on 2006 January 1-3 (Program ID: 076.D-0381(B)), 
by using the NTT 3.5m 
telescope and the near IR imager/spectrometer SOFI, equipped 
with a 1kx1k HAWAII array detector. 
All the observations have been performed by using  
0.292''/pixel scale, providing a $\sim$~5''x 5'' field of view each frame. 

Total integration times of 4min in J, 8min in H and 16min 
in $K_s$ (split into sets of shorter exposures) 
have been secured, allowing to reach a magnitude threshold of J$\sim$19 and 
H and $K_{s}\sim$18.5. All the secured images have been 
roughly centered on the cluster center. Moreover, for each target cluster, 
a control field (a few arcminutes away from each cluster center) 
has been observed adopting 
the same instrumental configuration; these field images have been used 
to construct median-average sky frames. High signal-to-noise 
flat fields in each band have been acquired by using a halogen lamp 
alternatively swichted on and off. The final cluster and field frames have 
been sky-subtracted and flat-field corrected.
The observations have been obtained in good seeing conditions 
(0.6''-0.7'' on average)
\footnote{Only for the cluster NGC 361 the observations have been performed in  
worse seeing conditions ($\sim$1'') limiting the  
magnitude threshold to $K_{s}\sim$18.}.
The point spread function fitting procedure has been performed 
by using the ALLSTAR routine of the DAOPHOT \citep{stet87} reduction package. 
The detection of the stellar sources has been performed in the J image, then 
this list of stellar objects has been used as reference for the reduction of the 
images in the other 
two filters. 
The output catalog, obtained by cross-correlating the single-filter catalogs, 
includes all stars measured in at least two bands. The instrumental 
magnitudes have been transformed into the Two-Micron All-Sky Survey (2MASS) 
photometric system, by using the large number of stars (a few hundreds) 
in common between SOFI and the 2MASS. 
No significant color term in each band has been found. Finally, 
the brightest stars that turned out to be saturated in the SOFI images 
have been recovered from 
the 2MASS catalog (the typical saturation limit 
in our images is
$K_{s}\sim$11).
A final catalog listing $\sim$1500-2500 stars has been obtained in each 
program cluster. 
Each cluster center of gravity 
$C_{grav}$ (see Table~\ref{tab1}) has been computed by averaging 
the $\alpha$- and $\delta$-coordinates of stars lying within a fixed radius 
(typically $\sim$90'') 
from a {\sl guess} center (estimated by eye).

\section{The CMDs}
\label{cmd}

\begin{figure}[h]
\plotone{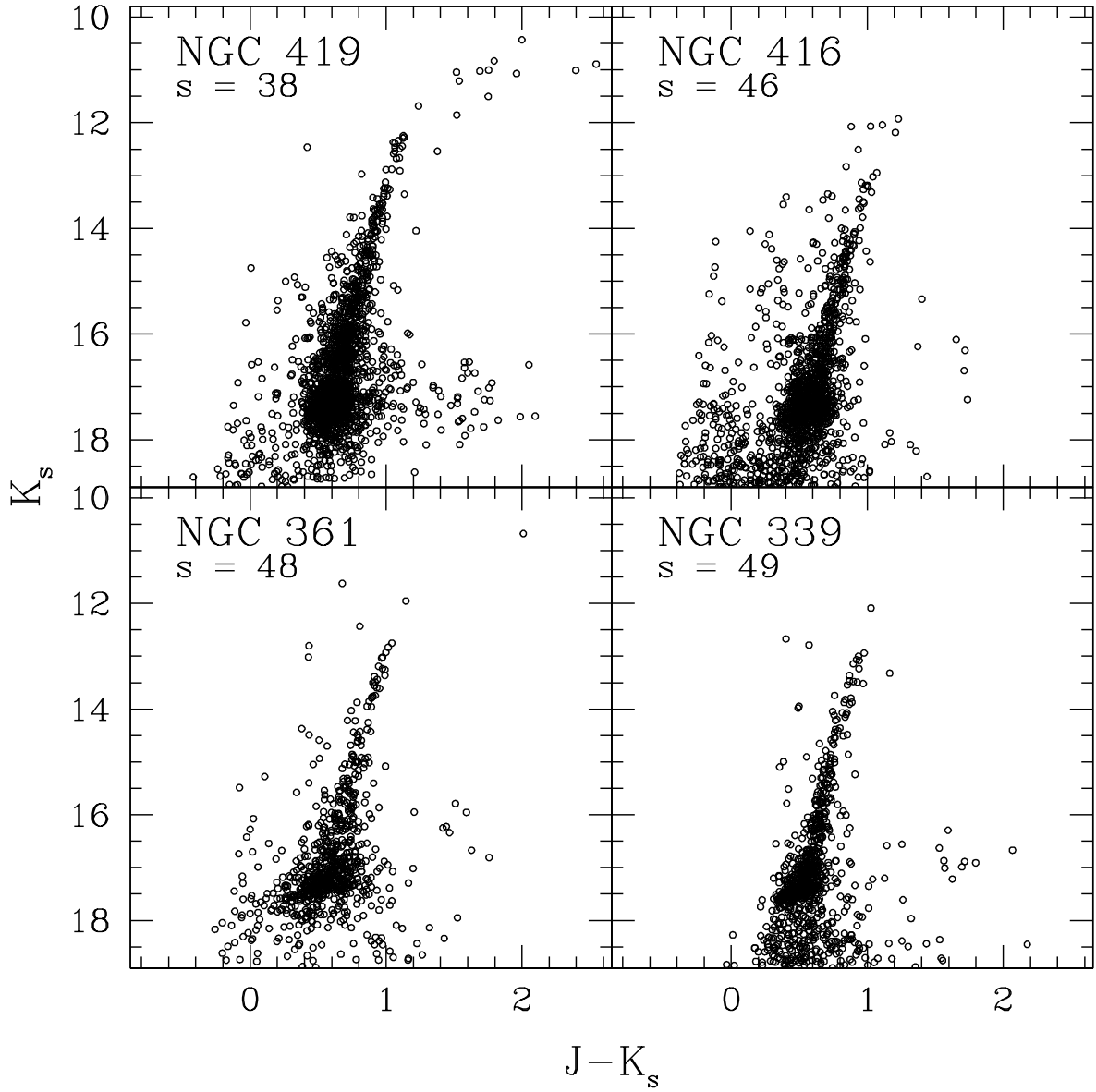}
\caption{Observed ($K_{s}$, J-$K_{s}$) CMDs of the four observed SMC clusters.}
\label{cmd1}
\end{figure}

\begin{figure}[h]
\plotone{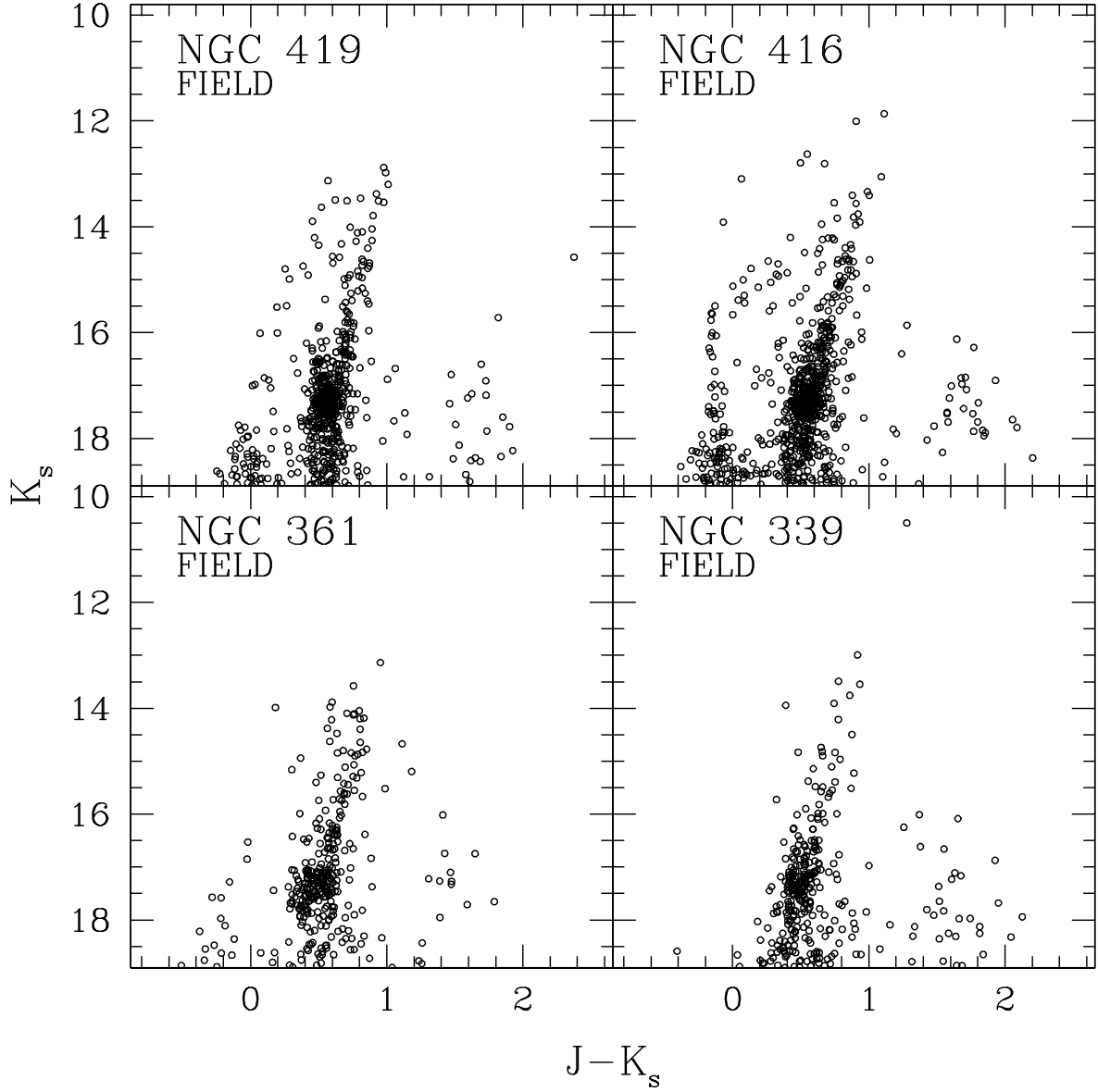}
\caption{Observed ($K_{s}$, J-$K_{s}$) CMDs of the fields adjacent to the four observed SMC clusters.}
\label{cmd2}
\end{figure}

Fig.~\ref{cmd1} reports the ($K_{s}$, (J-$K_{s}$)) 
CMDs of the four clusters 
of our sample. For each cluster the age s-parameter 
\footnote{The so-called s-parameter, defined by \citet{ef85} 
represents an extension of the SWB classification, introduced by \citet{swb} to date the 
Magellanic clusters and based 
on the position of the clusters 
in the $(U-B)_0$-$(B-V)_0$ plane.} \citep{ef85} is also reported; 
Fig.~\ref{cmd2} shows the CMDs of the corresponding control
fields.
For each cluster, Table~\ref{tab1} summarizes 
the coordinates of the 
center of gravity and other cluster properties, in particular: \\ 
{\it (i)~reddening}: for 3 stellar clusters (namely NGC~416, 339 and 361) 
we adopted the E(B-V) computed by \citet{mighell},  
from  optical WFPC2@HST CMDs. For NGC 419 
the typical reddening value of the SMC \citep{hunter} has been adopted. 
However, it is worth noticing that the small amount of reddening in the direction 
of these clusters has a negligible impact on the near IR photometry. \\ 
{\it (ii)~age}: 
as already discussed in Paper I and II, the lack of a 
homogeneous age scale for the Magellanic clusters based 
on the measurement of the Main Sequence Turn-Off 
represents a severe limitation.
Accordingly to Paper I and II, we adopted the \citet{ef85} s-parameters 
as an age indicator 
and the most recent 
age calibration derived  by \citet{gir95}: log(Age)=6.227+0.0733$\cdot$s. 
By using WFPC2@HST photometry, 
\citet{mighell} 
derived Turn-Off ages for NGC~339, 361 and 416 
in the 5--7 ($\pm$ 1.1--1.3) Gyr range (see Table~\ref{tab1}).
Recently, new determinations of the ages of NGC~339, 416 and 419 have been 
presented by \citet{glatt}, based on high-resolution ACS@HST 
photometry, deriving an age of 6 Gyr for the first 2 clusters.
In the case of NGC 419 the authors list only an age range 
(between 1.2 and 1.6 Gyr) because of the complex Turn-Off 
morphology. This age is also in agreement with the 
previous estimate \citep[see][]{rich00}. All the direct 
age estimates are consistent with those inferred from the 
s-parameter.

The main features of these CMDs are summarized as follows:\\ 
(1) an extended and  fully populated RGB;\\ 
(2) a bulk of stars at $K_{s}\sim$17.5, corresponding to the He-Clump;\\ 
(3) the brightest objects with 
$K_{s}<$13 are likely AGB stars;\\ 
(4) the cluster and field population have similar features.\\   
 
The presence of a well-defined and populated RGB in each cluster is 
in agreement with the relatively old age of the clusters which have 
already experienced 
the RGB phase transition (as discussed in Paper I and II).
We note that the CMDs of NGC~416 and of
its surrounding field show a blue stellar population 
(located at (J-$K_{s}$)$\sim$-0.1 and with stars brighter than 
$K_{s}\sim$~16.5).
This younger population has been also detected  
in the optical photometry presented by \citet{mighell}.

\section{Metallicity}
\label{met}

As well known the properties (morphology and position) of the RGB 
are sensitive function of the overall metallicity of the population. Hence, 
the presence of a well populated RGB in the CMDs shown in Fig.~\ref{cmd1} 
allows us to derive a photometric measure of the cluster metallicity. 
Recently, \citet{val04} have presented a calibration of a set of 
morphological RGB parameters (in the IR plane) in terms of the 
cluster metallicity for a sample of old Galactic globular 
clusters, by adopting both the \citet{cg97} iron 
metallicity scale and the global metallicity [M/H], as computed 
by taking into account the enhancement of the $\alpha$-elements.
Since the few available chemical information
for the SMC stellar population indicate solar scaled value 
of the [$\alpha$/Fe] abundance ratio 
\citep[see][]{hill97}, we use the [Fe/H] scale as reference. 
We used 
the entire set of RGB parameters defined by \citet{val04} 
in the IR planes ($K_s$, $J$-$K_s$) and (H, J-H), namely 
the $(J-K_{s})_0$  color at different absolute magnitudes 
$M_K$=(-3,-4,-5,-5.5), the $(J-H)_0$  color at 
$M_H$=(-3,-4,-5,-5.5), the $K_s$ absolute magnitude at fixed 
($(J-K_s)_0$)=~3, the H absolute magnitude at $(J-H)_0$=~3 and 
the slope of the RGB. 
All the photometric parameters have been measured along the RGB 
cluster mean ridge lines.
These fiducial ridge lines have been 
computed following the procedure described in \citet{fer99,val04}.
First, we selected (by eye) stars belonging to the RGB 
in order to exclude He-clump, AGB and field stars, 
then the second-order polynom has been fitted to the observed 
distribution.
The ridge line has been transformed into the absolute plane 
by adopting a distance modulus of 
$(m-M)_0$=18.99 \citep{cioni00}, the reddening listed in Table~\ref{tab1} 
and the extinction law by \citet{rl85}. Once the photometric 
parameters were measured, the various estimates of the cluster 
metallicity were computed 
from the equations listed in Appendix A of \citet{val04}. 
All these estimates  turn out to be consistent one to each other 
(with average dispersion of $\sim$0.1), so we 
assumed for each cluster metallicity the mean value 
(and reported in Table~\ref{tab2}), thus finding 
[Fe/H]=-1.18, -1.08, -0.99 and -0.96 dex for NGC~339, 361, 
416 and 419, respectively. \\ 
\citet{mighell} give a photometric estimate of the 
cluster metallicity by 
using the RGB slope in the optical plane 
(see Table~\ref{tab1}). We find  their metallicities 
to be $\sim$0.3 dex more metal-poor than ours; this discrepancy 
can be mainly ascribed to the different adopted metallicity scales. 
Indeed,  
the relations by \citet{val04} are calibrated on the 
\citet{cg97} scale, while the $slope_{RGB}$-[Fe/H] relation by 
\citet{mighell} is on the \citet{zw84} scale.

\subsection{Integrated Magnitudes}
\label{mag}

In order to compute the integrated magnitudes of the target clusters, 
we performed aperture photometry, with different aperture radii centered 
on the center of gravity.
A crucial step in this procedure is the correct decontamination from the field population. 
To do this we
performed an equivalent aperture photometry  
on each control field. The resulting field luminosity has been subtracted 
to the cluster luminosity. 
The instrumental integrated magnitude was then calibrated into the 2MASS 
system following the procedure described in Sect.~\ref{obs}.\\ 

For each cluster, Table~\ref{tab2} lists the integrated $K_{s}$ magnitude, 
(J-$K_{s}$) and (H-$K_{s}$) colors and the  
$K_{s}$ and bolometric luminosities, as computed 
by adopting an aperture radius representative of the entire cluster 
extension (typically 90-100'')
\footnote{In all the clusters the bulk of the luminosity lies within a  
$\sim$50-60'' radius}.

The formal error in the integrated magnitudes 
can be obtained as the quadrature sum of the photometric 
error associated to the task PHOT and the uncertainty in the aperture centering. 
The latter has been estimated by computing aperture photometry using 
4 different centers, shifted by $\pm$5 pixels with respect to 
the $C_{grav}$ coordinates. Typically, we estimated an error of $\sigma\sim$
0.04-0.05 mag in each band, that translates in a errorbar of $\sim$0.07 mag 
in color.\\ 
Total luminosities (in K$_s$-band and bolometric) have been 
computed by using the integrated magnitudes 
and adopting a distance modulus for the 
SMC of $(m-M)_0$=18.99 \citep{cioni00}, bolometric 
corrections (by using the $(J-K_{s})_0$ color) empirically calibrated by 
\citet{mon98} and solar values of 
$M^{Bol}_{\odot}=4.75$ and $M^{K}_{\odot}=3.41$. In the following all the 
derived luminosities are expressed in units of $10^4 L_{\odot}$. 
The main sources of error in this case are the 
uncertainty in the integrated magnitudes 
and in the bolometric corrections (an additional variation of $\sim$10\%)
which translate into a $\sim$5\% and $\sim$10\% uncertainty in luminosity.

\section{Star counts and population ratios}          
\label{count}

In order to estimate the RGB and AGB contributions 
to the total cluster light, we use star counts and population 
ratios, as obtained by adopting suitable {\sl selection boxes} for 
each evolutionary sequence (He-Clump, RGB and AGB), as discussed 
in Sect.~\ref{agb} and ~\ref{rgb} (see also Paper I and II). Two main effects must 
be taken into account in the definition of these quantities, the 
incompleteness of the photometric catalog and the contamination by 
field stars.

\subsection{Completeness and field decontamination}
\label{comp}

The degree of completeness can be quantified by adopting the
widely-used artificial star technique, discussed in
\citet{mateo88}.  
For each cluster we have
derived the RGB fiducial line and then a population of
artificial stars, having magnitudes, colors, and luminosity functions 
resembling the observed distributions has been generated and
added to the original images (by using the DAOPHOT task ADDSTAR). 
The frame area
sampling the cluster  has been
divided in 3 concentric regions with radii $r<$20"', 20''$\le r <$60'' and 60''$\le r <$90'', 
in order to take into account different crowding conditions 
and the completeness has been estimated independently in each of them. 
The maximum spatial extension of each cluster  has been estimated from
the cluster radial density profile.  
A total of $\sim$200,000
artificial stars   have been simulated  in each cluster  in
about 1000 simulation runs. Indeed, in order to not alter the crowding
conditions, $\sim$100-200 stars have been simulated in each run, 
corresponding to $\sim$10\% of the total stellar population.
The fraction of recovered objects in each magnitude
interval has been estimated  as
$\Lambda=\frac{N_{rec}}{N_{sim}}$: the
completeness curve was obtained in each radial 
subregion and shown in Fig.~\ref{comp}.
The correction for incompleteness in each radial region was 
performed by dividing each observed distribution by the corresponding
$\Lambda$ factor. The total number of stars has been obtained 
by summing the number of stars in each subregion.

It is worth noticing that this procedure allows to take 
into account only the loss of faint stars due to 
the crowding but not the possible excess of bright stars  due to
blending effects of two or more faint stars into a
brighter one. However, this latter effect is marginal in the 
near IR.

Another important effect which needs to be investigated, is
the degree  of contamination of the selected samples by the
foreground/background stars.  In this paper we have applied
a statistical decontamination, by using a control
field adjacent to the cluster.
The total number of stars observed in each evolutionary
sequence  (AGB, RGB and He-clump) has been counted
accordingly to the {\it selection boxes}  both in the 
cluster and field CMDs, and 
corrected for incompleteness (see above).  
The star counts in the field population have been
scaled to take into account the different surveyed area, 
and their contribution have been subtracted from the
cluster population. 

In summary, 
for each radial region, each selection box 
corresponding to each evolutionary stage has
been divided in bins of magnitude (typically 0.2 mag
wide). Then, the "corrected" number of stars in each bin
has been computed as follows:
$$ n_{corr} = n_{obs} + (n_{obs} (1/\Lambda -1)) - n_{f}$$
where $n_{obs}$  is the  number of stars observed in that bin,
the second term is the number of stars lost for incompleteness,
$n_{f}$  is the expected  number of  
field stars. 
The  total luminosity of each evolutionary stage can be computed
accordingly to the following relation:
$$ L_{corr} = (\sum^n_{i=1} L^{obs}_{i}) + (n_{comp}\times L_{eq}) -
(n_{f} \times L_{eq})$$
where the term $\sum^n_{i=1} L^{obs}_{i}$ is the total luminosity of 
stars observed in a given bin, $n_{comp}$ is the number of
stars lost for incompleteness, $n_{f}$  is the expected 
number of field stars, and $L_{eq}$ is the equivalent
luminosity of that bin, that is the luminosity of a star with
magnitude equal to the mean value of the bin.\\
Finally, star counts and total luminosity of each
evolutionary stage have been obtained by summing the  
contribution of all the bins.
\footnote{The typical errors for the different population ratios (by number and 
luminosities) have been estimated accordingly to 
the following formula  
$\sigma_{R}=\frac{\sqrt{R^{2}\cdot\sigma_{D}^{2}+\sigma_{N}^{2}}}{D}$  
with $R=N/D$, $N$ being the numerator and $D$ the denominator of the ratio, 
and by assuming that star counts follow a Poisson statistics.}

\section{The AGB and C-stars population}
\label{agb}
The AGB stars are the main 
contributors to the integrated SSP light between $\sim10^8$ and 
$\sim10^9$ yrs \citep{rb86,m98}.
AGB stars are initially Oxygen-rich, 
but if massive enough (with an initial mass M$>\sim$2$M_{\odot}$) 
a star undergoes the so-called
{\sl Third Dredge-Up} event
during the Thermal-Pulse (TP-AGB) phase,  
and freshly processed Carbon is carried to the surface,producing a C-rich AGB star.\\
\citet{fmb90} analyzed 
39 Magellanic clusters in order to identify AGB stars and concluded 
that up to 40\% of the bolometric luminosity comes from stars 
with $M_{bol}<$-3.6, likely belonging to the TP-AGB phase. 
In Paper II we studied the AGB population in the young-intermediate 
LMC clusters (with ages less than $\sim$3 Gyr), finding that the maximum 
contribution of the AGB to the cluster light occurs at an age of $\sim$500-700 
Myr, with a dominant contribution from the C-stars population.

We classify  AGB stars only those stars that satisfy the 3 following {\sl criteria}:\\ 
(1) stars brighter than ${(K_{s})}_{0}$=12.62 
\citep[corresponding to the RGB tip level for the SMC, see][]{cioni00}, 
in order to minimize the impact of possible RGB stars contamination;\\  
(2) in order 
to separate the cluster stars and the most bright background/foreground 
stars, only stars locate in the box showed in Fig.~\ref{ccd} are considered; \\ 
(3) stars located within the cluster extension 
(see Sec.~\ref{mag}).

In order to accurately select C-stars in each cluster, we adopted as 
a diagnostic tool the $(J-H)_0$-$(H-K_{s})_0$ color-color diagram. 
Left panel of Fig.~\ref{ccd} shows the cumulative $(J-H)_0$-$(H-K_{s})_0$ 
diagram for all the 
selected AGB stars, with overplotted the box defined from \citet{bessel} 
to isolate the mean locus of the C stars and long period variables (LPV), 
and the mean locus for the K giant stars defined by \citet{fpam78}.
All the stars flagged as C-stars show very red (J-$K_{s}$) colors also 
in the 2MASS database, thus excluding major 
errors in the PSF-fitting procedure.
The adopted methodology to identify and distinguish C and O-stars is also 
consistent with the criterion adopted by \citet{ch03}, 
who based their selection on the J-$K_s$ color only and 
targeting C-stars does brighter than the RGB-Tip and with (J-$K_s$)$>$1.3.

Right panel of Fig.~\ref{ccd} shows the cumulative CMD 
for the four clusters of the sample with 
overplotted the box used for the selection: the candidate AGB 
stars are marked (grey points for the O-rich and black points for the 
C-rich stars). 
In the following we summaries the census of the AGB stars in 
each cluster:\\
\begin{itemize}
\item {\sl NGC~416---}    
Only 1 AGB (O-rich) star has been identified in this cluster. 
The other bright stars detected in the field of view  
are located too far away from the cluster center (at $>$100''), 
hence they are unlikely cluster members. 

\item {\sl NGC~419---}
This cluster exhibits a large population of candidate AGB stars, 
already studied by \citet{fmb90} who identified 10 C-stars  
likely members. 
We found 21 AGB stars, 12 C-stars, 
4 likely LPVs,
within $\sim$90'' from the cluster center. 
None of the LMC clusters previously studied 
(Paper II) shows a comparable number 
of C-stars. The paucity of C-stars in the observed LMC clusters 
with the same age could 
be ascribed to a metallicity effect \citep[see Fig. 5 in][]{m05}.
We note that outside the cluster radius only 3 O-rich AGB stars 
have been detected and none C-stars.

\item {\sl NGC~339---}
One O-rich 
AGB star has been detected, just above the RGB Tip.

\item {\sl NGC~361---}
Only one AGB (O-rich) star has been identified. 

\end{itemize}

A possible source of error in the computation of 
the AGB counts and luminosities
is the location of the RGB Tip: 
however, a variation of 0.2 mag implies the inclusion of a few 
fainter AGB stars only (with a variation in the total AGB 
luminosity $<$10\%).

Table~\ref{tab3} lists the final star counts and luminosities of the AGB and C stars 
in each cluster. 
Fig.~\ref{agb1} (top panel) shows 
the $K_s$-band luminosity of the AGB stars normalized to the total 
luminosity as a function of age (black points);  
for comparison the values obtained for the LMC clusters 
in Paper II are also plotted (grey points).  
It is worth noticing the higher ($\sim$50\%) 
luminosity ratio of 
NGC 419 compared to the value ($\sim$10\%) of the other 3, significantly older clusters.
Fig.~\ref{agb1} (bottom panel) shows the same distribution but binned 
in age as discussed in Paper II for what concerns the LMC clusters, while 
only three out of four SMC clusters, in our sample namely NGC 339, 361 and 416, with similar ages 
have been binned.

Theoretical predictions computed by \citet{m98,m05} for [M/H]=~--0.33 
(solid line) and ~--1.35 dex (dashed line)
are also plotted for comparison.  
The theoretical AGB and RGB population ratios
have been computed by using SSP models by \citet{m98} and \citet{m05}, 
obtained with an evolutionary code that estimates the energetics of any 
post-main-sequence stage by following the prescriptions of the 
{\sl fuel consumption theorem} defined by \citet{rb86} (see also \citet{f04} for 
more details). 

Similarly, Fig.~\ref{agb2} shows the 
$K_s$-band luminosity of the C-stars only, normalized to the total
luminosity, as a function of age.
Note that in NGC~419 $\sim$80\%
of the AGB light is provided by the C-stars population. 

We thus confirm 
previous results discussed 
by \citet{fmb90} that C-stars are only detectable in relatively young ($<$2 Gyr) 
clusters of IV-VI SWB Type 
and theoretical predictions which require a minimum envelope mass 
for the occurrence of the {\sl Third Dredge-Up}. 
Stars with 
$<$1.2 $M_{\odot}$ initial mass have a residual (if any)  envelope mass which is 
too small to experience the {\sl Third Dredge-Up} .

\section{The RGB population}                         
\label{rgb}

In order to calculate the RGB population ratios we 
adopted the same procedure used in Paper I and II. 
Three observables have been identified to study the degree of 
development of the RGB as a function of age and metallicity: 
{\sl (i)}~
the number of RGB stars normalized 
to the number of He-Clump stars ($N_{RGB}$/$N_{He-Cl}$), 
{\sl (ii)}~
the bolometric luminosity of the RGB normalized to the He-Clump one 
($L_{RGB}^{bol}$/$L_{He-Cl}^{bol}$), and 
{\sl (iii)}~
the bolometric luminosity 
of the RGB normalized to the total cluster luminosity  
($L_{RGB}^{bol}$/$L_{TOT}^{bol}$). 
In order to identify the mean loci of the upper RGB and He-Clump stars, 
we use the cumulative, dereddened $K_0$-$(J-K)_0$ CMD 
as a diagnostic diagram.
As in Paper I and II (see their Fig. 5 and Fig. 9 
respectively)
we define two boxes for these evolutionary stages. 
The size of each box has been defined to sample the bulk 
of the population, assuming to be $\sim$ 5 times the photometric 
uncertainty at a given level of magnitude. The upper limit of the RGB box is 
the magnitude of the RGB Tip, the same used to define the bottom 
limit of the AGB box 
(Sect.~\ref{agb}).

The final population ratios (by counts and luminosities) have been computed 
following the procedure described in Sect.~\ref{comp}, by also applying 
the incompleteness 
correction and the statistical field decontamination. 
The results 
(star counts for the He-Clump and bright RGB and the corresponding 
bolometric luminosities) are reported in Table~\ref{tab4}. 
In NGC~419 
star counts and luminosities have been computed excluding the innermost 
region (with a radius of 20''), where  
completeness at the He-Clump magnitude level drops down to 60\% 
(as shown in Fig.~\ref{comp}).

Fig.~\ref{rgb1} and \ref{rgb2} plot the resulting observables for 
the four SMC clusters (black points) 
presented in this study and for the LMC clusters (grey points) discussed 
in Paper I and II. 
Theoretical predictions 
for the  [M/H]=~--0.33 and ~--1.35 dex (solid and dashed line respectively) 
metallicities are also plotted 
for comparison. 

The cluster NGC~419 displays population ratios (both in counts and luminosities) 
slightly higher with respect to the theoretical predictions 
(similarly to the LMC cluster NGC~1783 in Paper II) 
but still consistent with the occurrence of the RGB phase transition.
NGC~339, 361 and 416
show $N_{RGB}$/$N_{He-Cl}$ ratios
somewhat in between the two model predictions. 
Their location is consistent 
with our photometric estimates of the cluster 
metallicity (see Table.~\ref{tab2}), 
slightly higher than the ones obtained by \citet{mighell}. 

\section{Conclusions}
\label{conc}

By using high-quality near IR photometry of four  
SMC stellar clusters with intermediate ages we derived 
new photometric metallicities.
All the observed clusters show similar metallicities 
([Fe/H]$\sim$-1) with only a weak dependence with the age. 
Actually, the Age-Metallicity Relation for the SMC is not well 
known and the different metallicity indicators, namely stellar clusters, 
planetary nebulae, 
field stars, still exhibit strong discrepancies.
The most recent survey of SMC field giants 
based on Ca II triplet by 
\citet{carrera} indicates a value of [Fe/H]$\sim$--1 dex 
in the age range between $\sim$3 and $\sim$10 Gyr, 
and [Fe/H]$\sim$--0.7 dex at $\sim$1 Gyr. 

Furthermore, we investigated the contribution of the AGB and 
RGB evolutionary stages to the total cluster luminosity.
The cluster NGC~419, with an age of $\sim$1 Gyr, exhibits 
population ratios for AGB and RGB that follow the behaviour already 
observed in the LMC clusters for objects of similar age.
The other 3 clusters, with older ages in the  $\sim$5-7 Gyr range 
show a negligible ($\sim$6\%) luminosity contribution by the AGB, 
lacking  bright C-stars, and an increasing contribution by the RGB 
population with respect 
to clusters of younger ages like NGC~419 and those in the LMC. 
We find a general agreement between the empirical population 
ratios and those predicted by theoretical models at [M/H]=~--1.35 dex. 

\acknowledgements  
We warmly thank the anonymous referee for his/her 
useful suggestions.
This research 
was supported by the  
Ministero dell'Istruzione, del\-l'Uni\-versit\`a e della Ricerca 
and it is part of the Progetti Strategici 2006 granted by the 
University of Bologna

\begin{deluxetable}{cccccllc} 
\tablecolumns{8} 
\tablewidth{0pc}  
\tablecaption{Main parameters of the sample of observed SMC clusters}
\tablehead{ 
\colhead{Cluster} &    \colhead{$\alpha$(J2000)}& 
\colhead{$\delta$(J2000)}  & \colhead{s}& 
\colhead{$Age^s$}
 &  \colhead{$Age^{TO}$}
 &  \colhead{[Fe/H]} &   \colhead{E(B-V)}\\
  &   & & &(Gyr) & (Gyr)  &  & }
\startdata 
\hline
NGC~419   & 01:08:17.35 & -72:53:04.30 & 38 & 1.0 & $1.2-1.6^e$  & $-0.60^b$                       &	$0.08^d$    \\
NGC~416   & 01:07:58.82 & -72:21:18.96 & 46 & 4.0 & $5.6^a$;~$6^e$  & $-0.80^b$;$-1.44^a$             &	$0.08^a$ \\
NGC~361   & 01:02:10.09 & -71:36:18.73 & 48 & 5.6 & $6.8^a$  & $-1.45^a$                       &	$0.07^a$ \\
NGC~339   & 00:57:46.19 & -74:28:17.58 & 49 & 6.6 & $5.0^a$;~$6^e$   & $-0.70^b$;$-1.12^c$  ; $-1.50^a$&	$0.03^a$ \\
\hline
\enddata 
\tablecomments{$~~~~~$\\
Units of right ascension are hours, minutes and seconds, and units of declination are 
degrees, arcminutes, and arcseconds. The s-parameter is from \citet{ef85}.
$Age_s$ and $Age_{TO}$ indicate the ages inferred by the calibration of s-parameter by 
\citet{gir95} and by direct TO measurements, respectively.
 The ages are from
(a)  \citet{mighell} and (e) \citet{glatt}. The metallicities are from (a)
\citet{mighell}, (b) \citet{defr}, (c) \citet{dh98}. 
The reddening are from (a) \citet{mighell} and (d) \citet{hunter}.}
\label{tab1}
\end{deluxetable}

\begin{deluxetable}{ccccllc} 
\tablecolumns{7} 
\tablewidth{0pc}  
\tablecaption{Integrated $K_s$ magnitude, colors, luminosities and corresponding 
metallicities of the target clusters}
\tablehead{ 
\colhead{Cluster} &    \colhead{$K_s$}& 
\colhead{(H-$K_s$)}  & \colhead{(J-$K_s$)}& 
\colhead{$L^k_{tot}$}
 &  \colhead{$L^{bol}_{tot}$}& [Fe/H] \\
  & & & & ($10^{4}L^{k_{s}}_{\odot}$)& ($10^{4}L^{bol}_{\odot}$)}
\startdata 
\hline
NGC~419   & 7.49 & 0.31 & 1.09 & 92.05& 23.22   & -0.96 \\
NGC~416   & 8.68 & 0.13 & 0.68 & 30.76& 14.01   & -0.99 \\
NGC~361   & 8.94 & 0.11 & 0.78 & 24.21&  9.73   & -1.08  \\
NGC~339   & 9.38 & 0.10 & 0.72 & 16.14&  6.84   & -1.18  \\
\hline
\enddata 
\tablecomments{The derived photometric metallicities are calibrated in the 
\citet{cg97} metallicity scale.}
\label{tab2}
\end{deluxetable}

\begin{deluxetable}{clllcccc} 
\tablecolumns{8} 
\tablewidth{0pc}  
\tablecaption{Star counts and luminosities for AGB and C stars}
\tablehead{ 
\colhead{Cluster} &    \colhead{$N_{AGB}^{obs}$} &
 \colhead{$N_{AGB}^{field}$} & 
 \colhead{$N_{Carb}$} &  \colhead{$L^{k}_{AGB}$/$L^{k}_{tot}$}
  &  \colhead{$N_{Carb}$/$L^{bol}_{tot}$}
 &  \colhead{$N_{Carb}$/$L^{k}_{tot}$} 
 &  \colhead{$L^{k}_{Carb}$/$L^{k}_{tot}$} }
\startdata 
\hline
NGC~419   & 21 & 0 & 12 & 0.54& 0.53 & 0.13& 0.38    \\
NGC~416   & 1  & 0 & 0  & 0.04& 0    & 0& 0      \\
NGC~361   & 1  & 0 & 0  & 0.06& 0    & 0& 0      \\
NGC~339   & 1  & 0 & 0  & 0.08& 0    & 0& 0      \\
\hline
\enddata 
\label{tab3}
\end{deluxetable}

\begin{deluxetable}{clllcccc} 
\tablecolumns{5} 
\tablewidth{0pc}  
\tablecaption{Star counts and luminosities for RGB and He-Clump stars}
\tablehead{ 
\colhead{Cluster} &    \colhead{$N_{RGB}$} &
 \colhead{$N_{He-Clump}$} & 
   \colhead{$L^{bol}_{RGB}$} 
 &  \colhead{$L^{bol}_{He-Clump}$}\\ 
  & & &($10^{4}L_{\odot}^{bol}$)& ($10^{4}L_{\odot}^{bol}$) }
\startdata 
\hline
NGC~419   & 263& 608 &8.69 & 3.85 	\\
NGC~416   & 190& 299 &5.47 & 1.66      \\
NGC~361   & 109 & 180 &4.25 & 1.23      \\
NGC~339   & 107 & 188 &3.09 & 1.20     \\
\hline
\enddata 
\label{tab4}
\end{deluxetable}

\begin{figure}[h]
\plotone{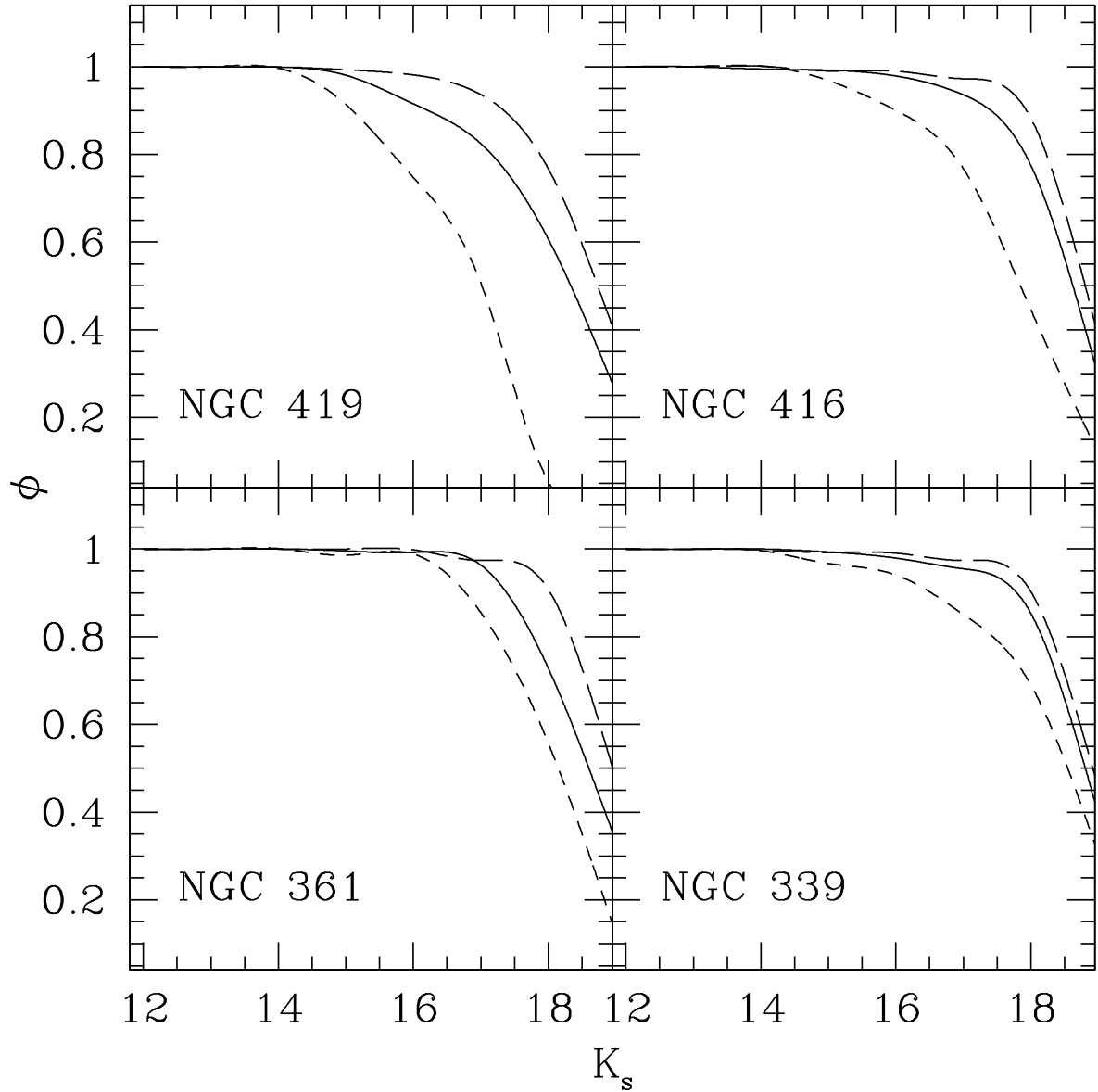}
\caption{Completeness curves for the four SMC clusters.
Short-dashed curves represent the inner region (r$<$20''), solid lines 
the region between 20 and $\sim$60'', and long-dashed lines the region 
between $\sim$60 and 90''.}
\label{comp}
\end{figure}

\begin{figure}[h]
\plotone{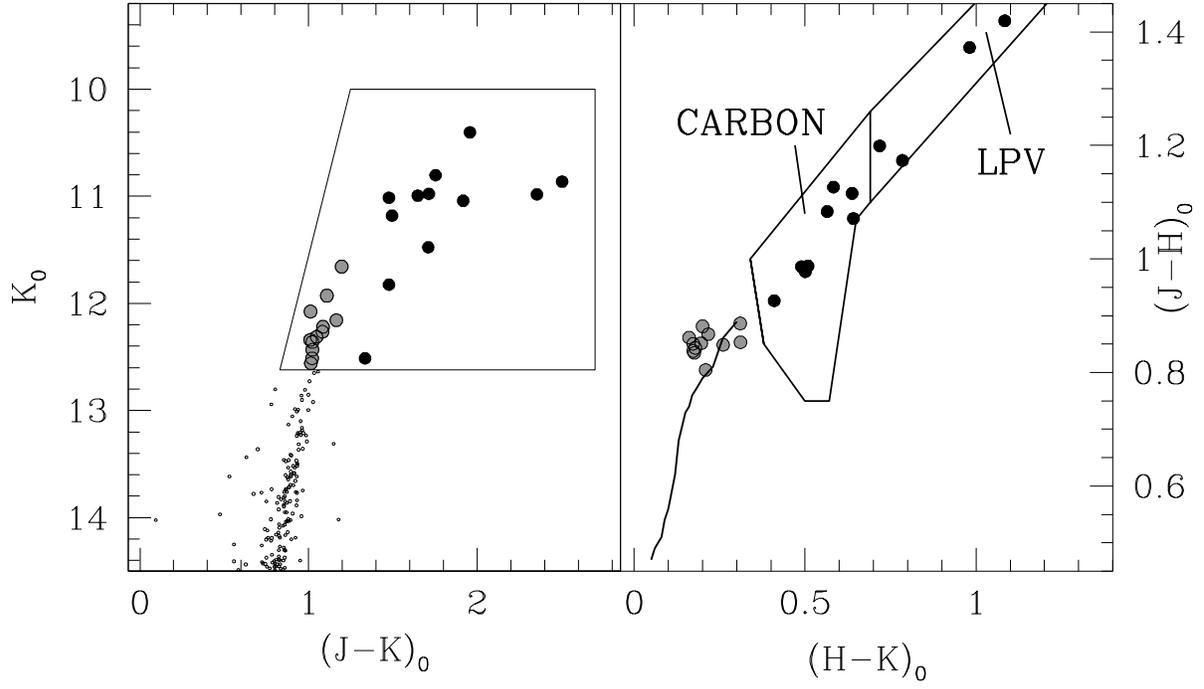}
\caption{Cumulative, dereddened ($K_0$, (J-$K_0$)) CMDs for the entire cluster sample, 
with overplotted the box used to select the AGB population. 
Grey points indicate the O-rich AGB stars and the black points the C-rich AGB stars. }
\label{ccd}
\end{figure}

\begin{figure}[h]
\plotone{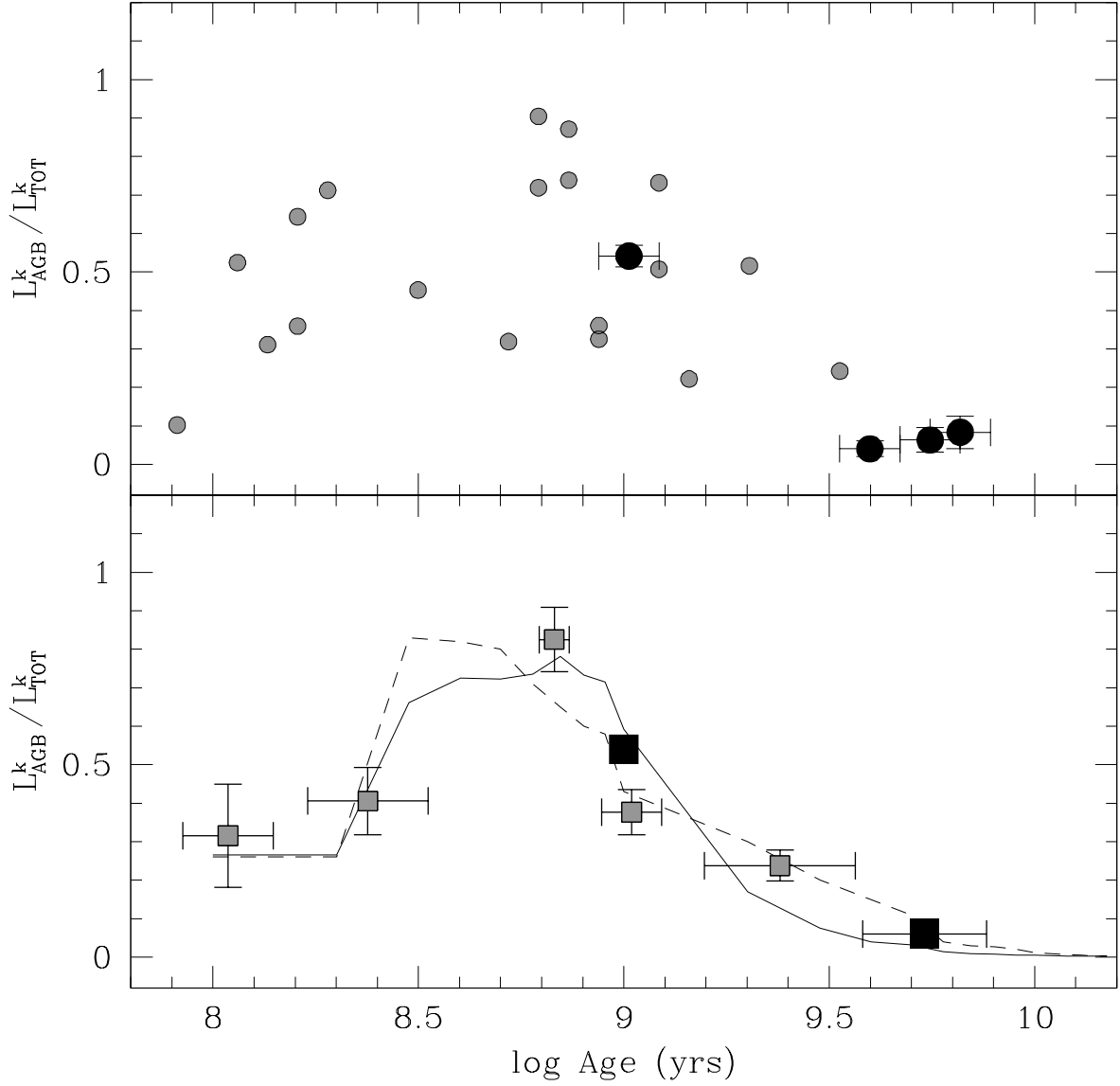}
\caption{Top: Observed AGB contribution to the total cluster $K_s$-band 
luminosity as a function of the age. The black points indicate the 
SMC clusters. The grey points indicate the LMC clusters studied in 
Paper II. Bottom: Mean and standard deviation of the same ratio with the clusters 
grouped into age-bins. Theoretical predictions for the temporal evolution of the entire 
AGB (both Early and Thermal-Pulse AGB) at different metallicities are overplotted: solid line 
indicates the model computed at [M/H]=~--0.33 and dashed line at [M/H]=~--1.35.}
\label{agb1}
\end{figure}

\begin{figure}[h]
\plotone{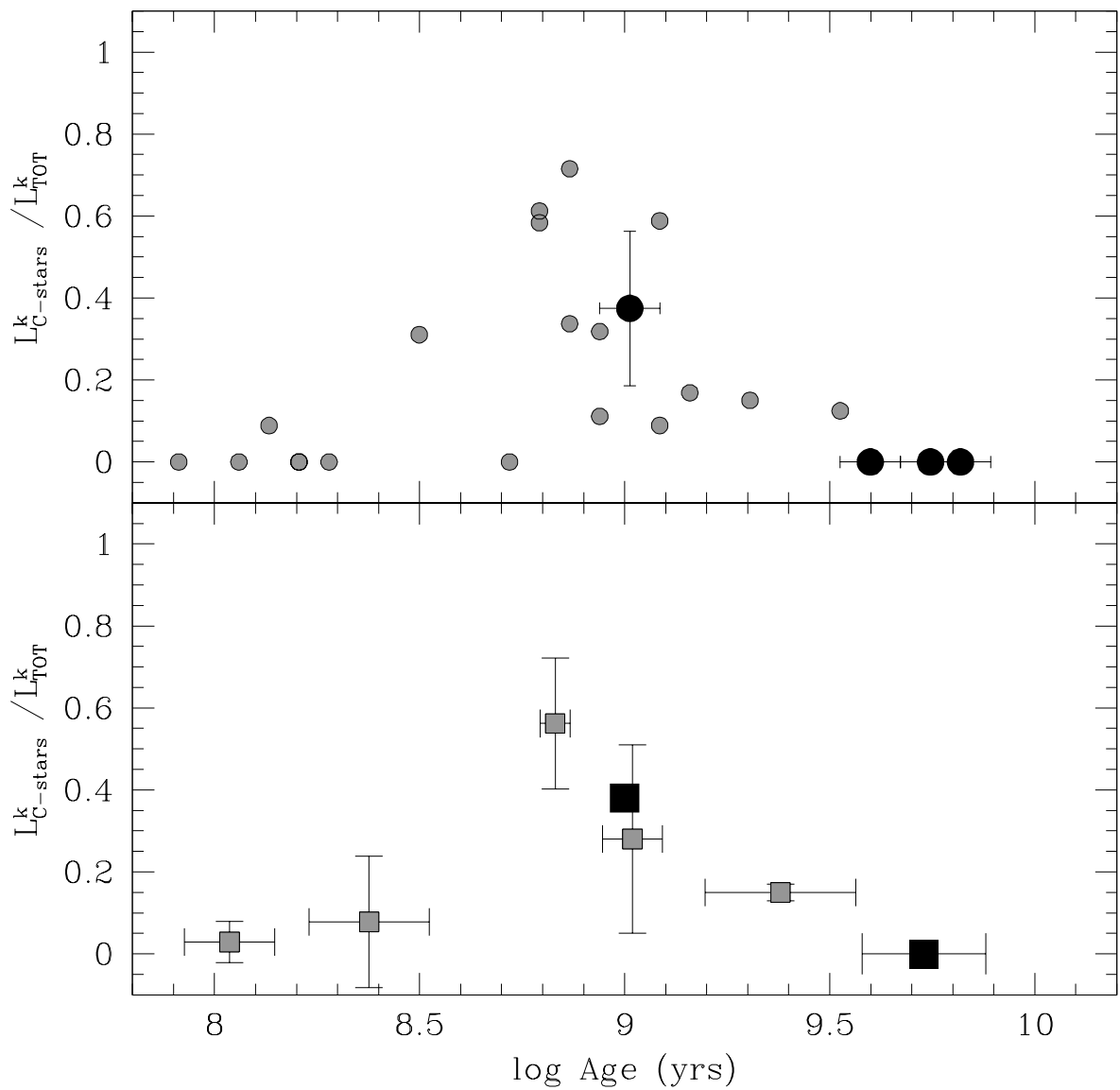}
\caption{Top: Observed C-star contribution to the total cluster $K_s$-band 
luminosity as a function of the age. The black points indicate the 
SMC clusters. The grey points indicate the LMC clusters studied in 
Paper II. Bottom: Mean and standard deviation of the same ratio with the clusters 
grouped into age-bins.}
\label{agb2}
\end{figure}

\begin{figure}[h]
\plotone{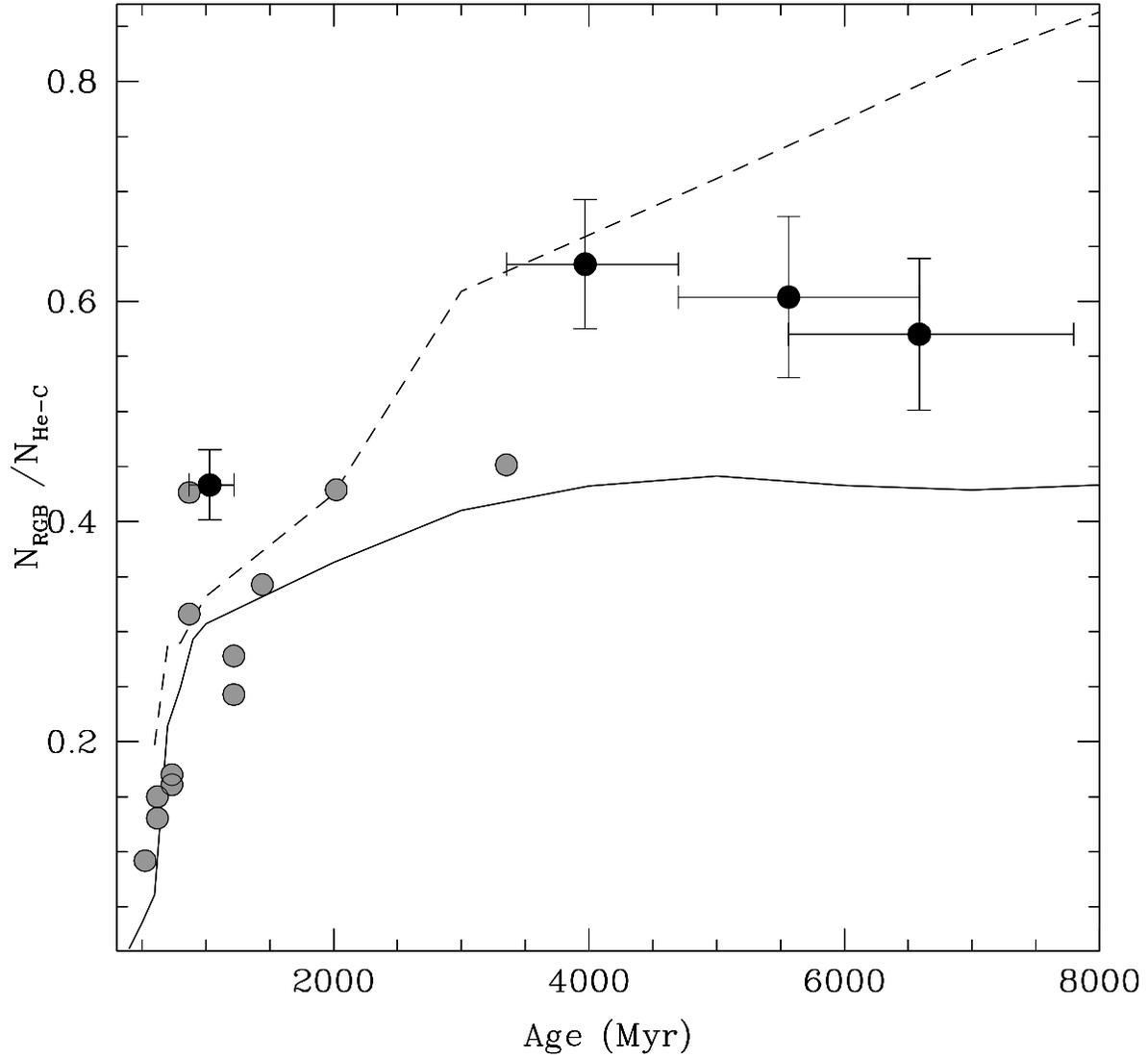}
\caption{Behaviour of the number counts of RGB stars normalized to 
the He-clump stars as a function of the age. Same symbols of Fig.~\ref{agb1}. 
Solid and 
dashed lines represent the theoretical predictions computed by using 
canonical models and global metallicity of [M/H]=-0.33 (solid line) and 
-1.35 (dashed line).}
\label{rgb1}
\end{figure}

\begin{figure}[h]
\plotone{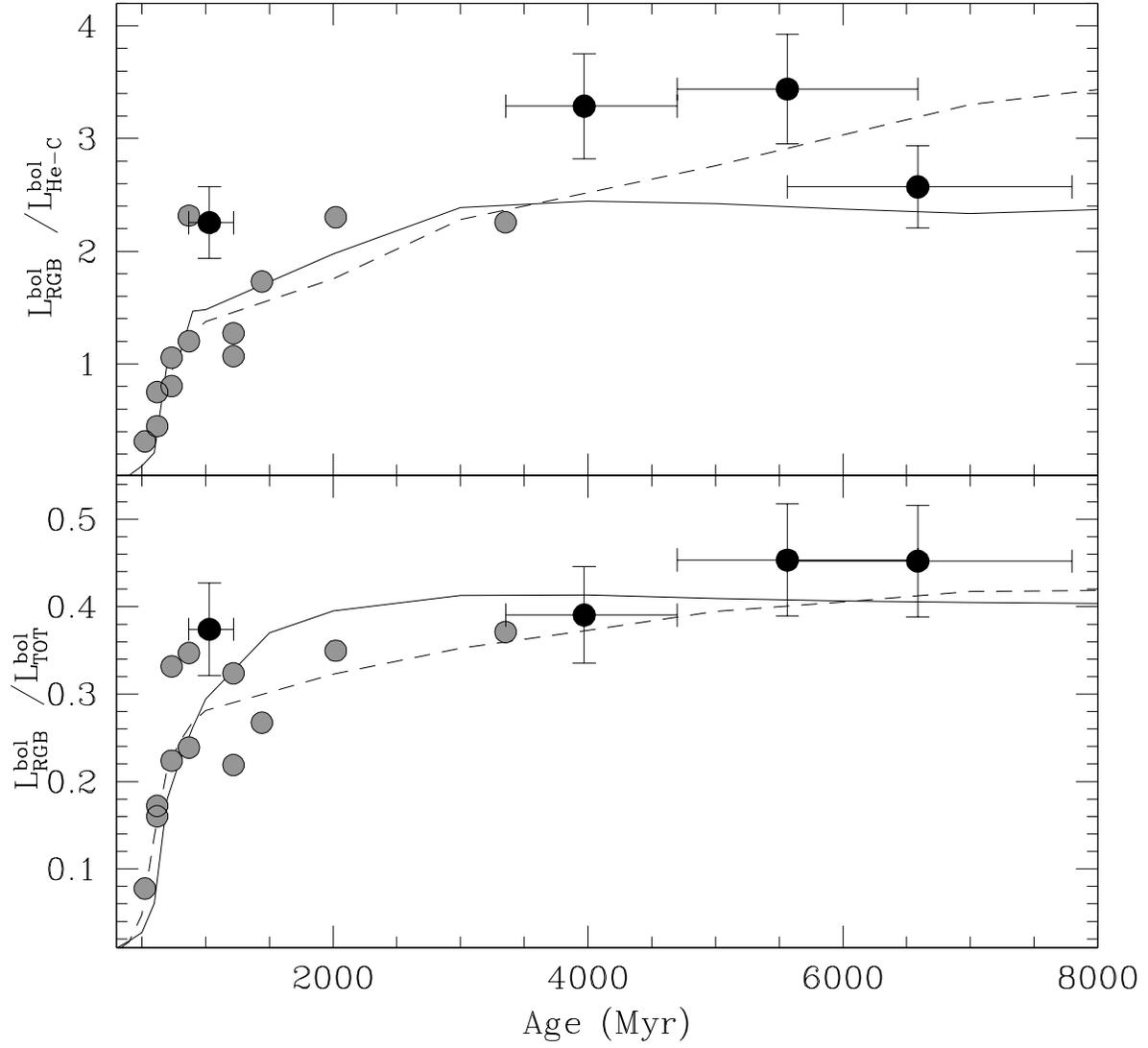}
\caption{Top: Bolometric luminosity of the RGB normalized to the 
He-clump as a function of the age for the observed MC clusters. Same symbols of Fig.~\ref{agb1}.
Bottom: Bolometric luminosity of the RGB normalized to the total 
luminosity for the same clusters. }
\label{rgb2}
\end{figure}

\end{document}